# Evolution of Orbits of Trans-Neptunian Bodies at the 2 : 3 Resonance with Neptune

## S. I. Ipatov* and J. Henrard**

*Keldysh Institute of Applied Mathematics, Miusskaya pl. 4, Moscow 125047, Russia

**Departement de Mathamatique, Faculte Universitaires Notre–Dame de Paris, Rempart de la Vierge 8, B-5000, Namur, Belgique

**Abstract**—The results of the numerical investigations of the evolution of orbits of trans-Neptunian bodies at the 2 : 3 resonance with Neptune are presented. The gravitational influence of the four giant planets was taken into account. For identical initial values of the semimajor axes, eccentricities, and inclinations, but for different initial orbital orientations and initial positions in orbits, we obtained different types of variations in the difference $\Delta\Omega = \Omega - \Omega_N$ in the ascending-node longitudes of the body and Neptune, and in the perihelion argument $\omega$. When $\Delta\Omega$ decreases and $\omega$ increases during evolution, then most of the bodies leave the resonance in 20 Myr. In the case of an increase in $\Delta\Omega$ and a decrease in $\omega$, the bodies stay in the resonance for a much longer time. Regions of eccentricities and inclinations, for which some bodies were in the $\eta_{18}$ secular resonance ($\Delta\Omega \approx$ const) and in the Kozai resonance ($\omega \approx$ const), were obtained to be larger than those predicted for small variations in the critical angle. Some bodies can at the same time be in both these resonances.

## INTRODUCTION

89 trans-Neptunian objects were known at the end of 1998. The semimajor axes $a$ of the orbits of 88 of them are in the interval from 35 to 49 AU. The diameters of these objects are between 100 and 400 km, and their characteristic masses are $m \sim 5 \times 10^{-12} M_{\odot}$, where $M_{\odot}$ is the mass of the Sun. The stellar magnitudes of the objects are ~22–24. The trans-Neptunian belt is often called the Kuiper belt or the Edgeworth–Kuiper belt in the honor of the scientists, who for the first time, predicted its existence. Now the number of objects in the trans-Neptunian belt with masses on the order of the masses of the observed objects ($d > 100$ km) is considered equal to 70000 (Jewitt *et al.*, 1996; Levison and Duncan, 1997). Jewitt *et al.* (1996) estimated the total mass of the present belt as $M_\Sigma \sim (0.06-0.25)m_\oplus$ at $a \sim 30-50$ AU. These authors considered that the actual width of the belt, which is proportional to the mean inclination, must exceed the observed width by a factor of 3. Like Pluto, about 35% of the observed trans-Neptunian objects are in the 2 : 3 resonance with Neptune. They are called Plutinos. Jewitt *et al.* (1998) consider that, due to observational selection, this portion is overestimated, and that actually only 10–20% of the trans-Neptunian objects in the zone of 30–50 AU are Plutinos.

The results of the investigations of the orbital evolution of some bodies moving in the 2 : 3 resonance with Neptune are presented in the papers by Morbidelli *et al.* (1995), Malhotra (1995), Levison and Stern (1995), Morbidelli (1997), Gallardo and Ferrar–Mello (1998). Morbidelli (1997) considered three types of orbits located in the 2 : 3 resonance: regular, strongly chaotic, and weakly chaotic ones. Elements of weakly chaotic orbits can change regularly for a long time (up to 1 billion years and more), and then begin to change chaoticly. Morbidelli believes that weakly chaotic orbits are the main present supplier of bodies to the Neptune's orbit. During the age of the Solar System, about half of these bodies, including $4.5 \times 10^8$ short-period comets, left the neighborhood of the 2 : 3 resonance. Due to the overlap with the secular resonances and the Kozai resonance, the 2 : 3 resonance is stable only at a small amplitude of libration, $\sigma = -2\lambda_N + 3\lambda - \pi$, where $\lambda = \omega + \Omega + M$ is the mean longitude, $\pi = \omega + \Omega$ is the perihelion longitude, $\omega$ is the perihelion argument, $\Omega$ is the ascending-node longitude, and $M$ is the mean anomaly. The value for Neptune is marked by $N$.

Gallardo and Ferrar–Mello (1998) investigated the evolution of some orbits, which are in the 2 : 3 resonance with Neptune, and established that $\Delta\Omega = \Omega - \Omega_N$ (where $\Omega_N$ is the longitude of the ascending node of Neptune) librates (the $\nu_{18}$ secular resonance) for the eccentricities $e < 0.05$ and $e > 0.28$. At $0.05 < e < 0.3$, transitions between a libration and a circulation are possible. The perihelion argument $\omega$ decreases (i.e., moves clockwise) at $e < 0.15$ and increases at $e > 0.3$. For $e \sim 0.24$, the argument of perihelion librates around 90° (the Kozai resonance). When studying the evolution of the orbits of bodies in the asteroid belt, Kozai (1962) obtained that, for large inclinations, $\omega$ can librate around 90° or 270°, and therewith the variations in the eccentricity $e$ and the inclination $i$ can be large. The period of variations in $\Delta\pi = \pi - \pi_N$ in the trans-Neptunian belt can be less than 3 Myr. The simultaneous libration of $\Delta\pi$ and $\Delta\Omega$ (this resonance, $\nu_8 + \nu_{18}$, was predicted by Morbidelli et al. (1995)) can cause highly inclined and eccentric orbits. Thomas and Morbidelli (1996) noted that the Kozai resonance does not influence the motion of bodies in the Kuiper belt, but affects the evolution of long-period comets.

## COMPUTER RUNS

We investigated the orbital evolution of trans-Neptunian bodies by numerical integration of the six-body problem (the Sun, four giant planets, and a test body). In order to take into account the gravitational influence of planets, we used the symplex method, i.e., the RMVS2 algorithm (Regularized Mixed Variable Symplex Algorithm) from the Swift integration package worked out by Levison and Duncan (1994). This algorithm provides a speed of calculations (often at the same accuracy of calculation) by an order of magnitude greater than the methods of numerical integration, which were worked out earlier. The initial positions and velocities of the planets were taken from the test of the SWIFT integration package.

To evaluate the accuracy of calculations for some typical investigated orbits, we compared the results of the calculations obtained with the use of the RMVS2 integrator with those obtained with the BULSTO method by Bulirsh and Stoer (1966). For one variant of the calculations, the plots of time variations in the orbital elements obtained with the use of the BULSTO integrator are presented in Fig. 1a; the analogous plots obtained with the use of the RMVS2 integrator are presented in Fig. 1b. In particular, the results obtained show that for quasi-periodical variations in orbital elements, the limits of variations in semimajor axes during 1 Myr differed by less than 5% for these two integrators, and the differences between the limits of variations in eccentricities and inclinations were much smaller.

The considered time span $T$ of the integration was not less than 20 Myr. Examples of the time variations in the orbital elements for $T = 20$ Myr are presented in Figs. 2a–2h. For many runs, the initial value of the semimajor axis of an orbit is $a_0 = 39.3$ AU. For such $a_0$, the resonant value of $a$ was always reached during evolution. At present, this value equals 39.6 AU, but it slightly varies with the variation in the semimajor axis of Neptune.

## TYPES OF EVOLUTION

Depending on the character of the time variations in $\Delta\Omega = \Omega - \Omega_N$ and the argument of perihelion $\omega$, Ipatov and Henrard (1996, 1997) considered several types of variations: $ID$ (Fig. 2a), $DI$ (Fig. 2b), $LI$ (Fig. 2c), $II$ (Fig. 2g), and others. Here, the first letter refers to the variation in $\Delta\Omega$, and the second corresponds to the variation in $\omega$; the letter "$I$" marks an increase, the letter "$\Delta\Omega$" marks a decrease; "$L$" is presented for the case of libration, and "$S$" for relatively small variations (no more than 360° during 20 Myr). For example, the $IL$ type corresponds to the case when $\Delta\Omega$ increases and $\omega$ librates (Fig. 2h). In contrast to the $L$ type, the time variations for the $S$ type do not look like a sinusoid. For some runs, the types of variations in $\Delta\Omega$ and $\omega$ change with time (Figs. 2d, 2e). The letter $L$ in the first position refers to the circular resonance $\nu_{18}$, and that in the second position, to the Kozai resonance.

For some initial data, variations in the eccentricity $e$, the inclination $i$, and the semimajor axis $a$ were relatively small. For other initial data, they were much larger, and bodies left the resonance, some of them being ejected into hyperbolic orbits. Even variations in the initial orbital orientations and the initial positions in orbits can highly influence the limits and the character of the variations in $e$, $i$, and $a$. For example, at $a_0 = 39.3$ AU, $e_0 = 0.15$, and $i_0 = 5°$, we considered 14 runs with different values of $\Omega_0$, $\omega_0$, and $M_0$ (where $M$ is the mean anomaly, and starting values for a body are designated by zero), and obtained that a body left the resonance for eight runs, i.e., for more than half of the cases. Among these eight orbits, which became nonresonant in 20 Myr, there were six of the $DI$ type (Fig. 2b), one of the $SI$ type, and one of the $LI$ type (Fig. 2c). The $ID$ type was obtained for three resonant orbits (Fig. 2a), and for the other three resonant orbits the type changed during evolution: $SI \longrightarrow IL$, $SD \longrightarrow SI \longrightarrow ID \longrightarrow IS$, and $SS \longrightarrow SI \longrightarrow SS \longrightarrow SI \longrightarrow IL$ (Fig. 2d). Some bodies were in the Kozai resonance (i.e., $\omega \approx$ const) for several Myr. For the above series of 14 runs at $T = 20$ Myr, the maximum eccentricity was smaller than 0.2 for five runs and exceeded 0.3 for three runs. At $\Omega_0 = \omega_0 = 0$ and $M_0 = 30°$, a body was ejected into a hyperbolic orbit after 37 Myr.

At $\Omega_0 = \omega_0 = M_0 = 60°$, $e_0 = 0.15$, and $i_0 = 5°$, we carried out runs for different values of $a_0$. For $38.8 \le a_0 \le 39.7$ AU, a body moved for some time (therewith, for more than 20 Myr at $39.0 \le a_0 \le 39.5$ AU) in the 2 : 3 resonance with Neptune. At $a_0$ equal to 37.5, 40.5, 41, and 41.5 AU, we obtained a libration of $\Delta\Omega$ of about 180° (the $\nu_{18}$ resonance). The $ID$ type was obtained for $39.1 \le a_0 \le 39.3$ AU, the $DI$ type was obtained for $38.5 \le a_0 \le 38.9$ AU and $39.6 \le a_0 \le 39.9$ AU, and changes in types ($SI \longrightarrow DI$, $ID \longrightarrow SD$, and $DI \longrightarrow SD$) were found for 39.0, 39.4, and 39.5 AU. At $i_0 = 5°$, $a_0 = 39.3$ AU and $\Omega_0 = \omega_0 = M_0 = 60°$, for $e_0$ from 0 to 0.3 (with a step of 0.05), we obtained the following types: $SI \longrightarrow SD$, $LI \longrightarrow LD$, $LD$, $ID$, $ID$, $II$, and $ID$. At $e_0 = 0.15$ and $a_0 = 39.3$ AU, for the above values of $\Omega_0$, $\omega_0$, and $M_0$, we had the $ID$ type for $0° \le i_0 \le 15°$ and the $IL$, $ID$, and $II$ types or

(a)

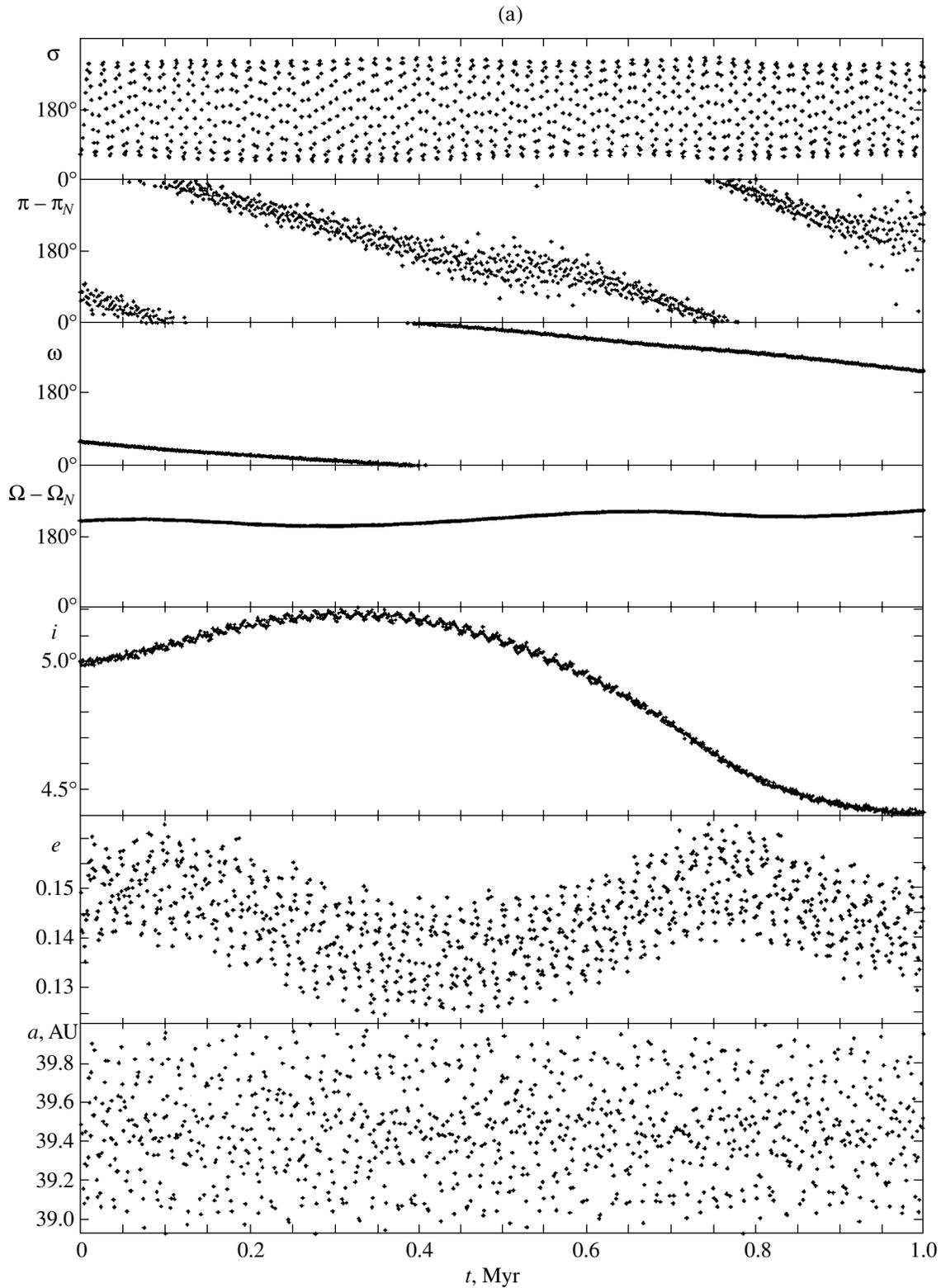

**Fig. 1.** Time variations in the semimajor axis $a$, eccentricity $e$, inclination $i$ of a body's orbit, the difference $\Omega - \Omega_N$ in the longitudes of the ascending node of the body and Neptune, the argument of perihelion $\omega$, the difference $\Delta\pi = \pi - \pi_N$ in the longitudes of the perihelion of the body and Neptune, and the critical angle $\sigma = 3\lambda - 2\lambda_N - \pi$, where $\lambda = \omega + \Omega + M$. Results are obtained by the numerical integration of equations of motion of the six-body problem (the Sun, the giant planets, and a body) with the use of the BULSTO integrator (a) and the RMVS2 integrator (b). Elements of the initial orbit of a body are: $a_0 = 39.3$ AU, $e_0 = 0.15$, $i_0 = 5°$, $\Omega_0 = \omega_0 = M_0 = 60°$.

(b)

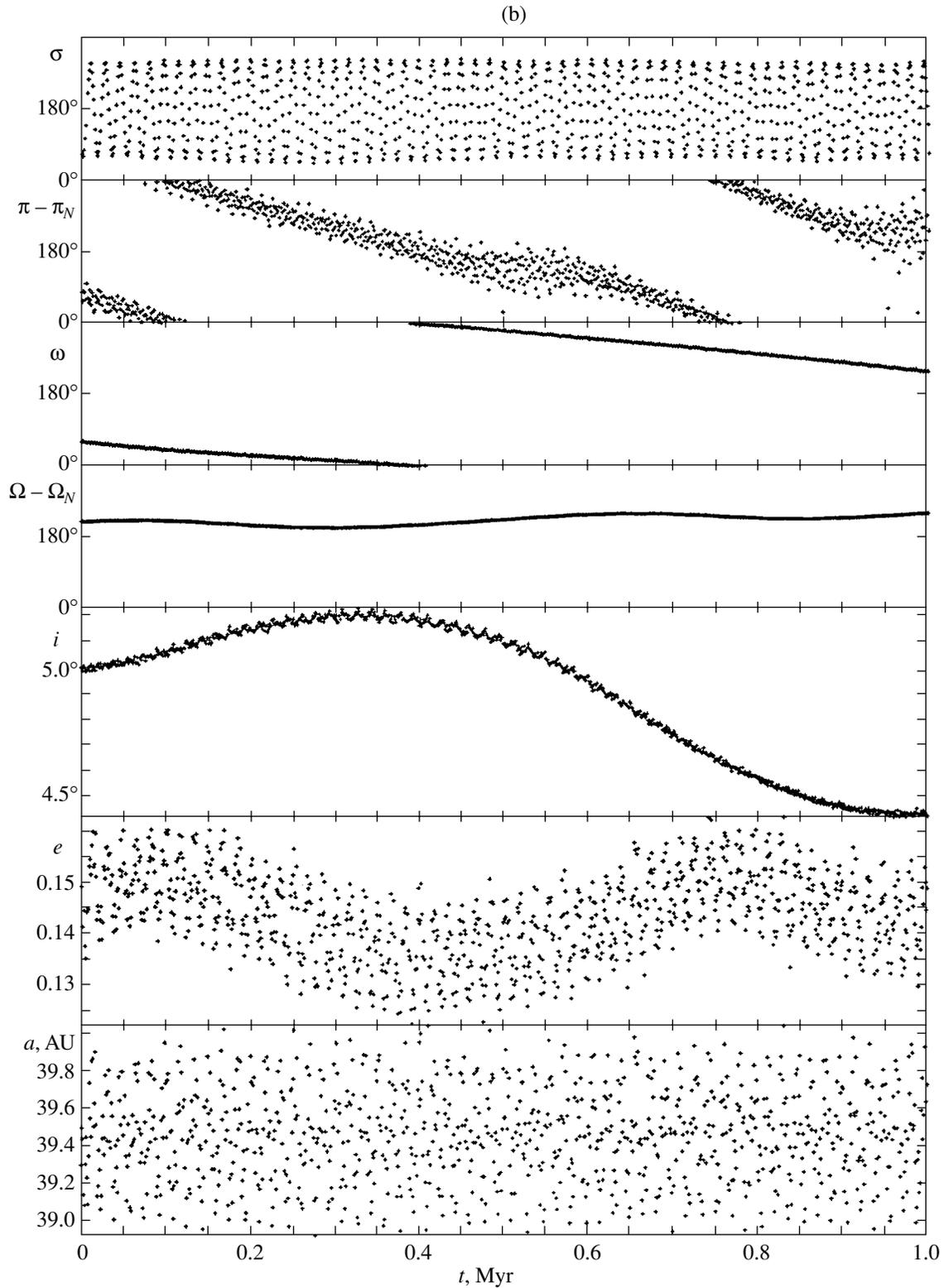

**Fig. 1.** Contd.

some combinations of these types for $30° \leq i_0 \leq 90°$. For most of the runs, variations in the critical angle $\sigma = 3\lambda - 2\lambda_N - \pi$ exceeded 180°. For nonresonant orbits we usually obtained the *DI*, *II*, or *LI* types. The *DD* and *DL* types were not obtained in our runs. For objects 1993SC, 1993SB, and 1993RO, the orbital evolution considered by Morbidelli *et al.* (1995) had the following types: *ID*, *II*, *ID* $\longrightarrow$ *IL* $\longrightarrow$ *II*, respectively.

(a)

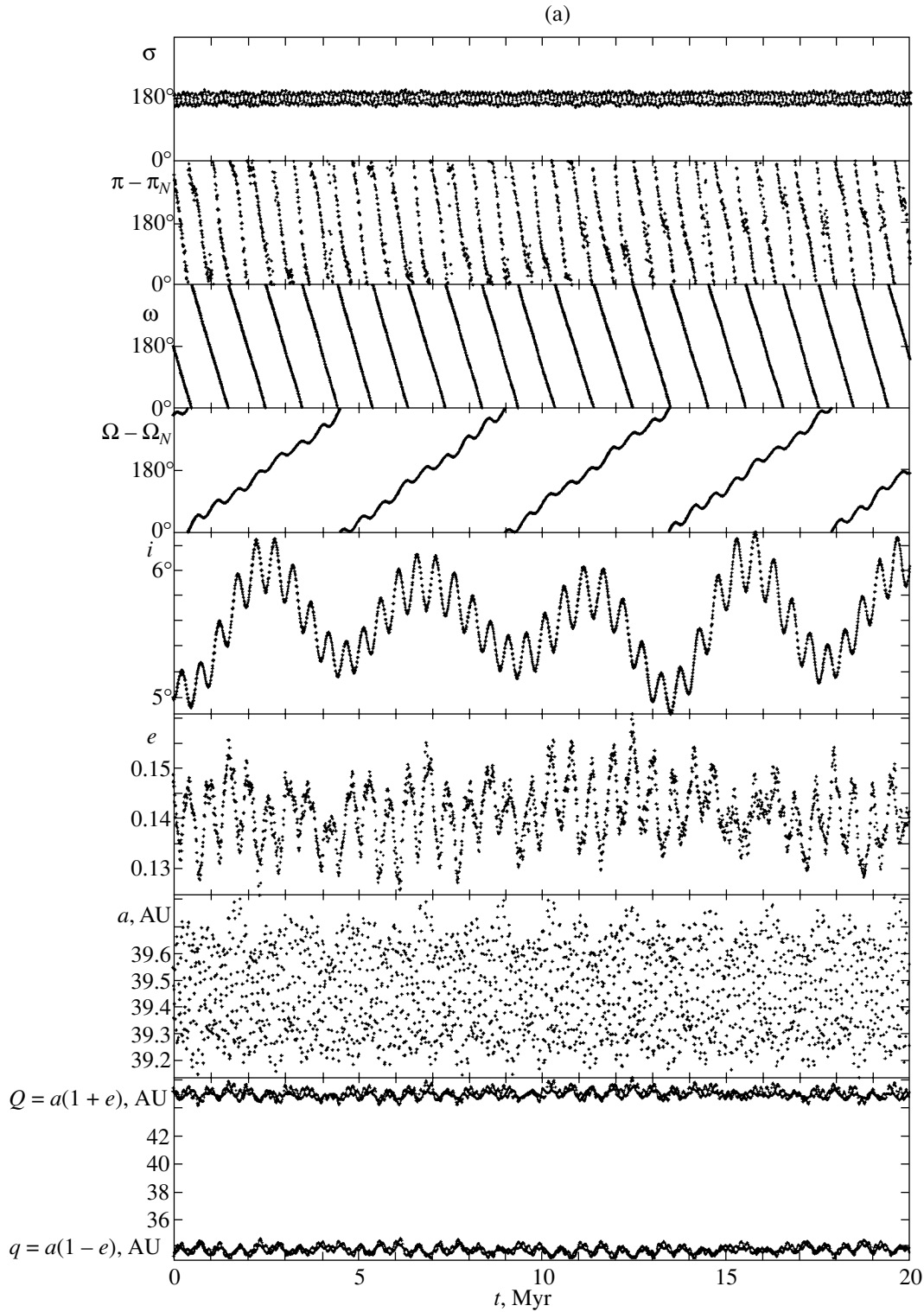

**Fig. 2.** Time variations (in Myr) in the semimajor axis $a$ (in AU), eccentricity $e$, aphelion and perihelion distances $Q = a(1 + e)$ and $q = a(1 - a)$ (in AU), inclination $i$ (in degrees), the difference $\Delta\Omega = \Omega - \Omega_N$ in the longitudes of the ascending nodes of the body and Neptune, the argument of perihelion $\omega$, the difference $\Delta\pi = \pi - \pi_N$ between the longitudes of the perihelion of the body and Neptune, and the critical angle $\sigma$ (all angles are given in degrees). For all variants, $a_0 = 39.3$ AU. Results were obtained with the use of the RMVS2 integrator. The influence of the giant planets was taken into account. (a) $e_0 = 0.15$, $i_0 = 5°$, $\Omega_0 = \omega_0 = M_0 = 180°$, $ID$ type; (b) $e_0 = 0.15$, $i_0 = 5°$, $\Omega_0 = 30°$, $\omega_0 = 0°$, $M_0 = 30°$, $DI$ type; (c) $e_0 = 0.15$, $i_0 = 5°$, $\Omega_0 = 60°$, $\omega_0 = 0°$, $M_0 = 60°$, $LI$ type; (d) $e_0 = 0.15$, $i_0 = 5°$, $\Omega_0 = 0°$, $\omega_0 = M_0 = 60°$, $SS \longrightarrow SI \longrightarrow SS \longrightarrow SI \longrightarrow IL$ types; (e) $e_0 = 0.05$, $i_0 = 5°$, $\Omega_0 = \omega_0 = M_0 = 60°$, $LL$ and $LI$ types; (f) $e_0 = 0.1$, $i_0 = 5°$, $\Omega_0 = \omega_0 = M_0 = 60°$, $LD$ type; (g) $e_0 = 0.3$, $i_0 = 10°$, $\Omega_0 = \omega_0 = M_0 = 60°$, $II$ type; (h) $e_0 = 0.3$, $i_0 = 45°$, $\Omega_0 = \omega_0 = M_0 = 60°$, $IL$ type.

(b)

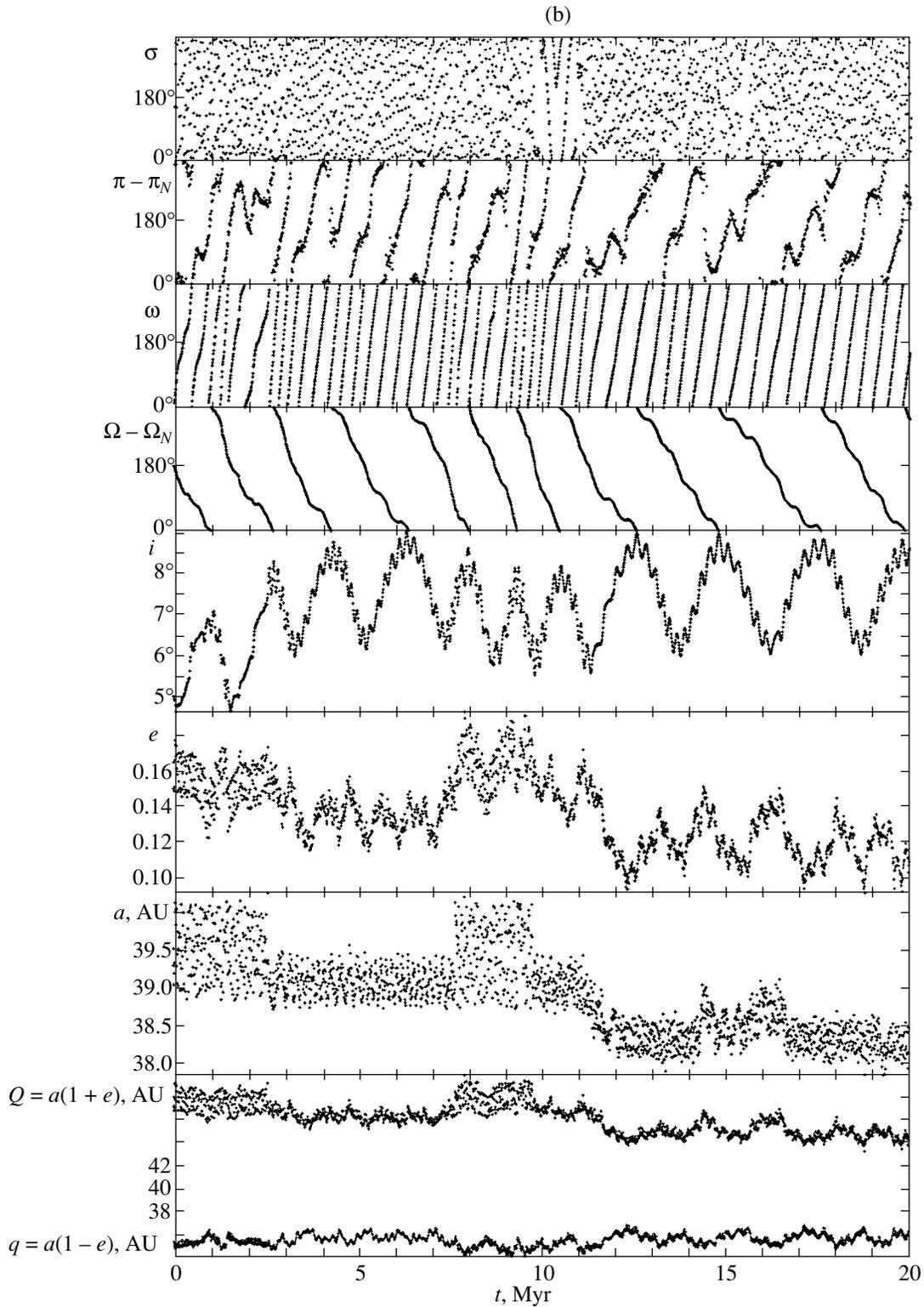

**Fig. 2.** Contd.

## VARIATIONS IN ORBITAL ELEMENTS

For orbits with a small amplitude of the $\sigma$ libration, the regions of $i$ and $e$ corresponding to the secular resonance $\eta_{18}$ ($\Delta\Omega \approx$ const) and the Kozai resonance are presented in Fig. 5 according to Morbidelli *et al.* (1995). These regions are far apart from each other: $e < 0.03$ and $i < 10°$ for the $\eta_{18}$ resonance, and $e > 0.2$ and $i < 10°$ for the Kozai resonance. In one of our runs, at $a_0 = 39.3$ AU, $e_0 = 0.05$ and $i_0 = 5°$, the values of $\Delta\Omega$ librated around

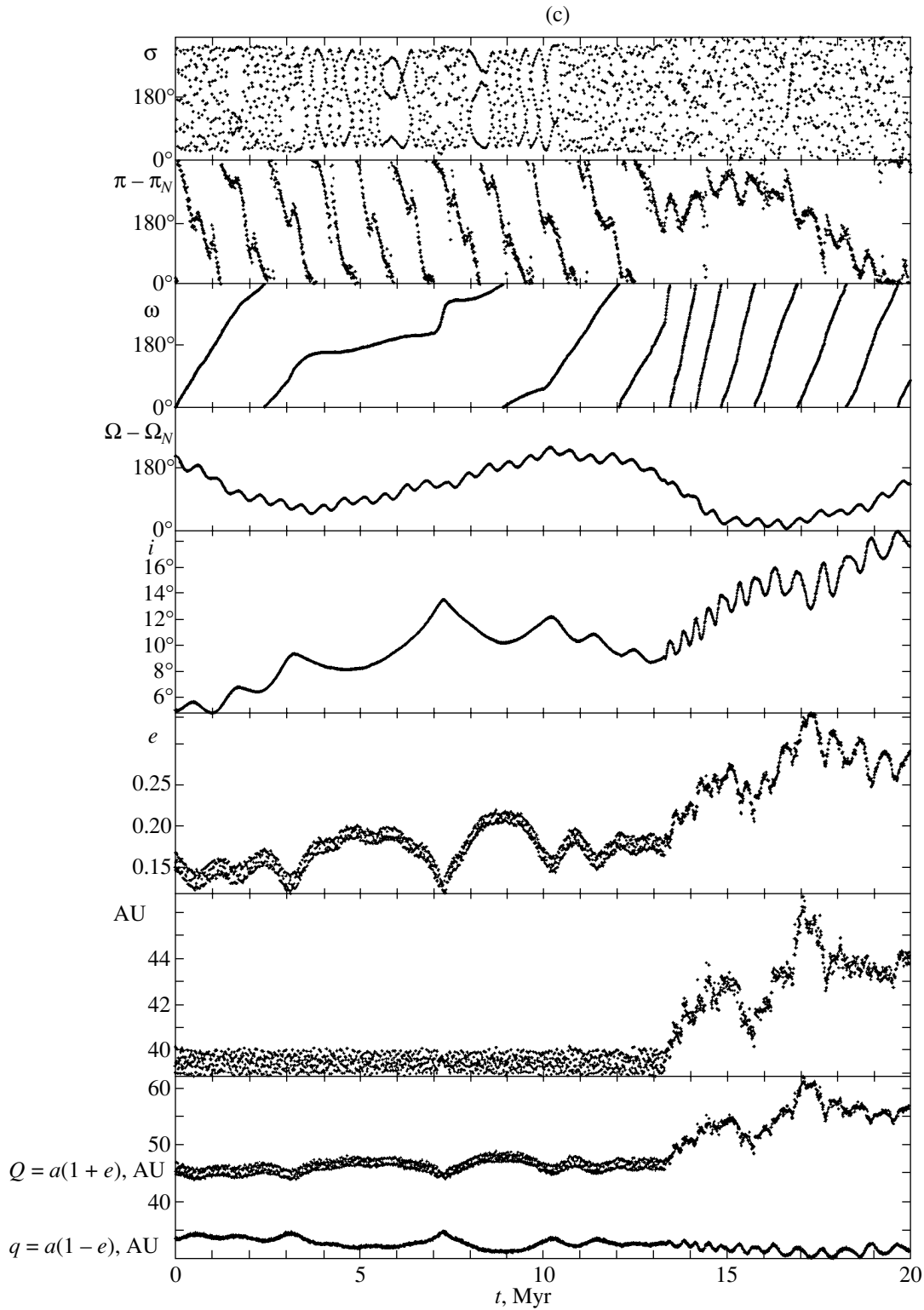

**Fig. 2.** Contd.

180° with an amplitude ~180°, and at the same time $\omega$ librated around 270° with an amplitude of ~100° during 6 Myr (Fig. 2e). The amplitude of the $\sigma$-libration was large (close to 360°) in this run. For other values of $\Omega_0$,

$\omega_0$, and $M_0$, at $e_0 = 0.05$ and $i_0 = 5°$, we usually obtained the *LI* type, but sometimes also the *DI* type. According to Fig. 5, in the work of Morbidelli *et al.* (1995), $\omega$ decreases at $e < 0.2$ and $i < 10°$ for a small amplitude of

(d)

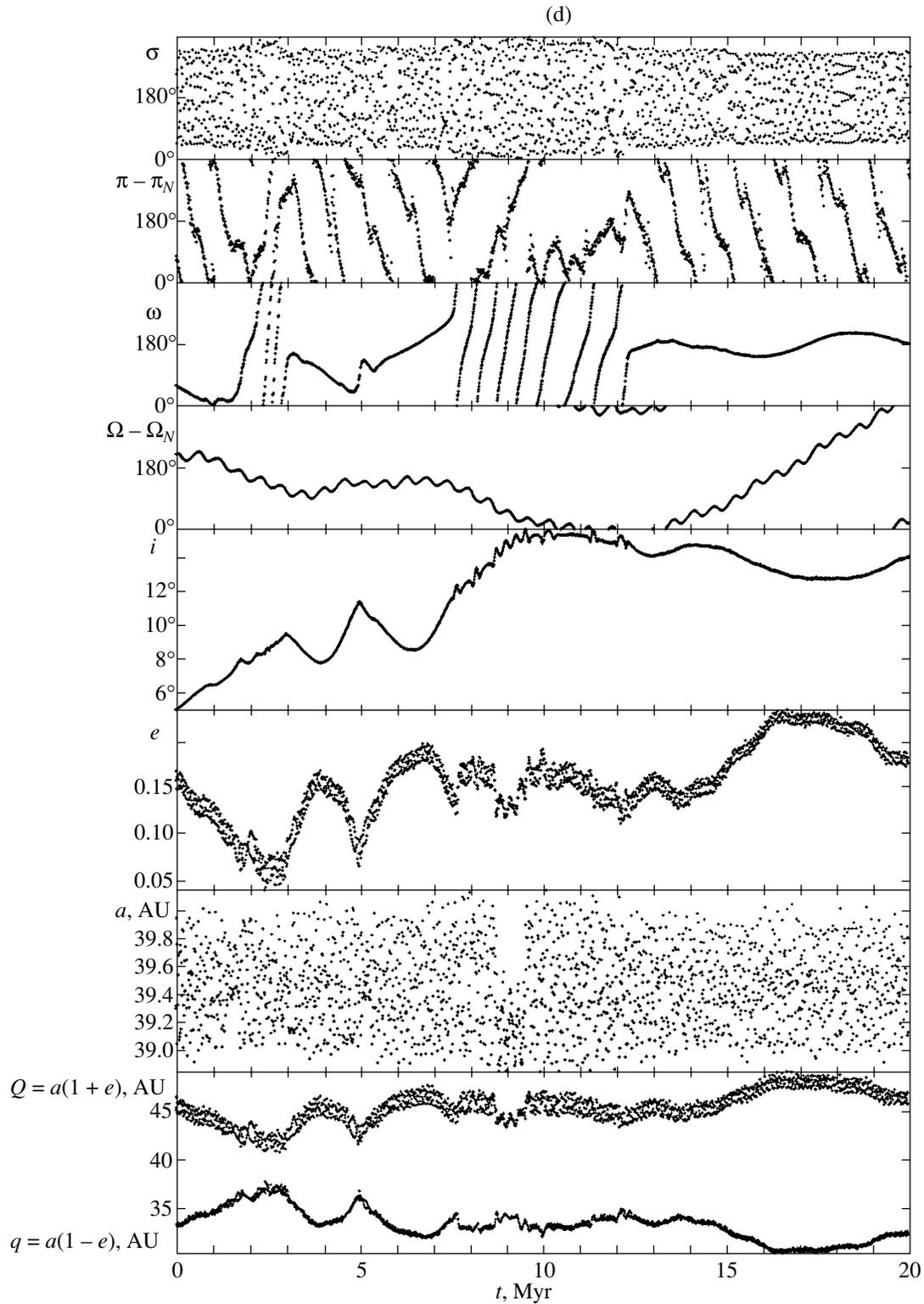

**Fig. 2.** Contd.

the σ-libration. For large variations in σ, we often obtained orbits with increasing ω for at these values of $e$ and $i$ (for example, for $e = 0.15$ and $i = 5°$). The region of values of $e$ and $i$ in our runs, for which the $\eta_{18}$ resonance was obtained for some orientations of orbits, was much larger than that in this figure. This resonance was obtained for most runs for $e_0 = 0.05$ and $e_0 = 0.1$.

For many runs, $i$ varies quasi-periodically with time, with a period equal to several million years, and $\Delta\Omega$

(e)

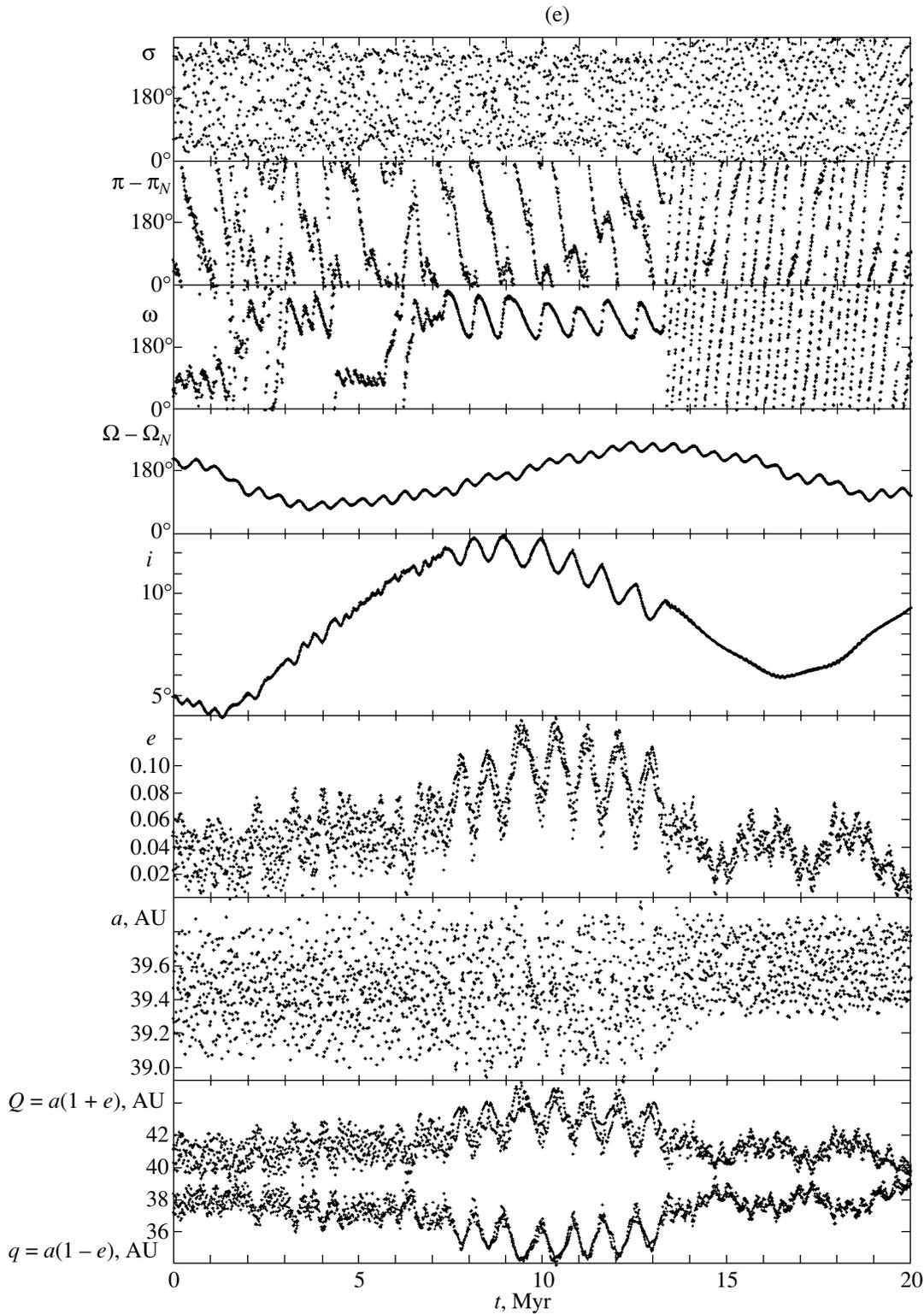

**Fig. 2.** Contd.

changes by 360° during this period. In this case, if $\Delta\Omega$ decreases in the course of evolution, then $\Delta\Omega = 0$ when $i$ reaches a maximum value, and $\Delta\Omega = 180°$ when $i$ reaches a minimum value (Fig. 2b). If $\Delta\Omega$ increases, then the maximum and minimum values of $i$ are reached at $\Delta\Omega$ equal to 180° and 0°, respectively (Fig. 2a). In some runs $\Delta\Omega$ librates, the width of the range of variations in $\Delta\Omega$ exceeds 180°, and this range

(f)

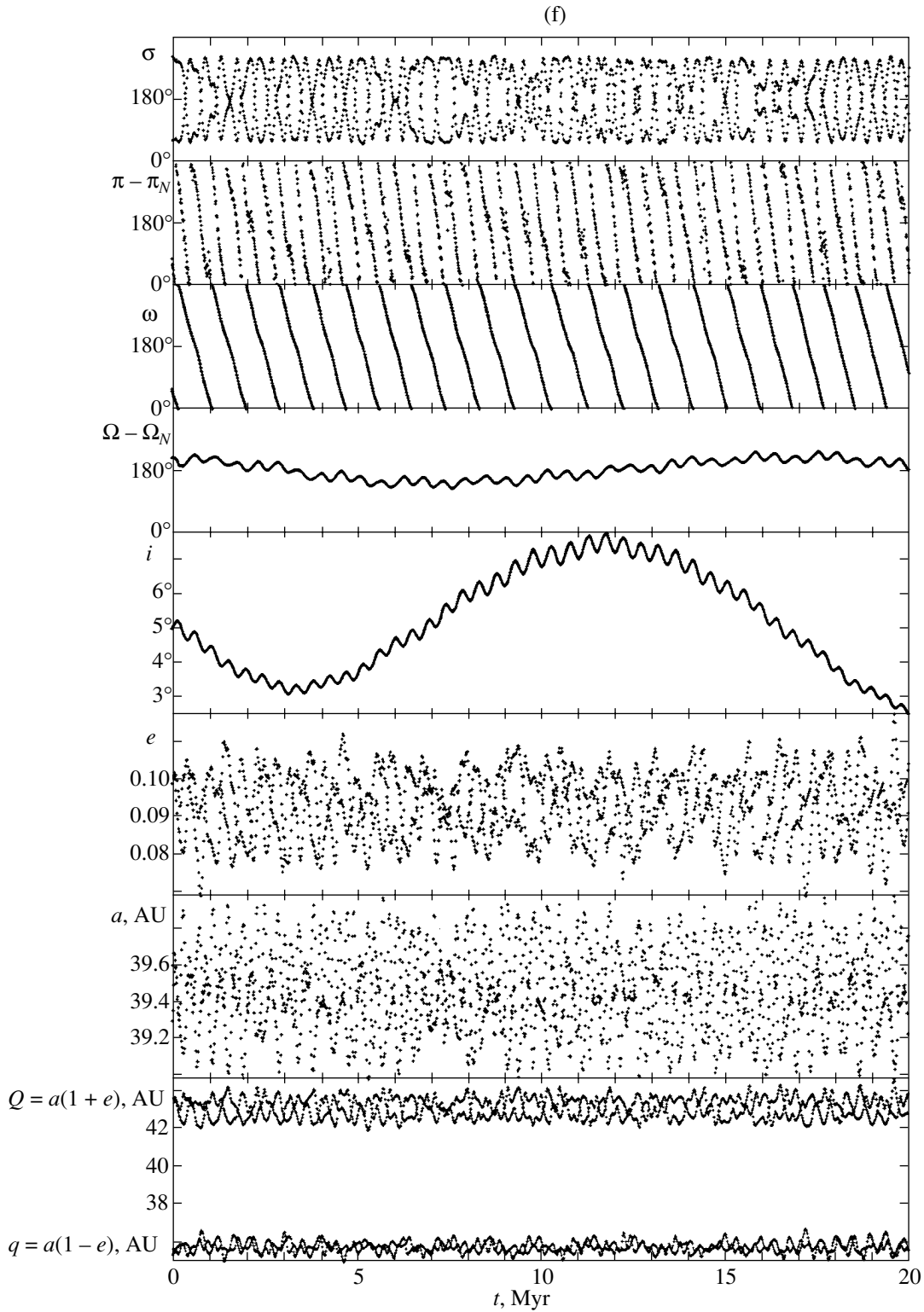

**Fig. 2.** Contd.

includes the value of $\Delta\Omega = 180°$. For example, $\Delta\Omega$ varies between 0° and 250°, between 70° and 260°, and between 90° and 250° in the runs presented in Figs. 2c, 2e, 2f, respectively.

In many runs we obtained variations in $e$ and $i$ with a period equal to $T_\omega/2$, where $T_\omega$ is the time interval during which $\omega$ decreases or increases by 360° (in the case of the Kozai resonance this is the period of libration).

(h)

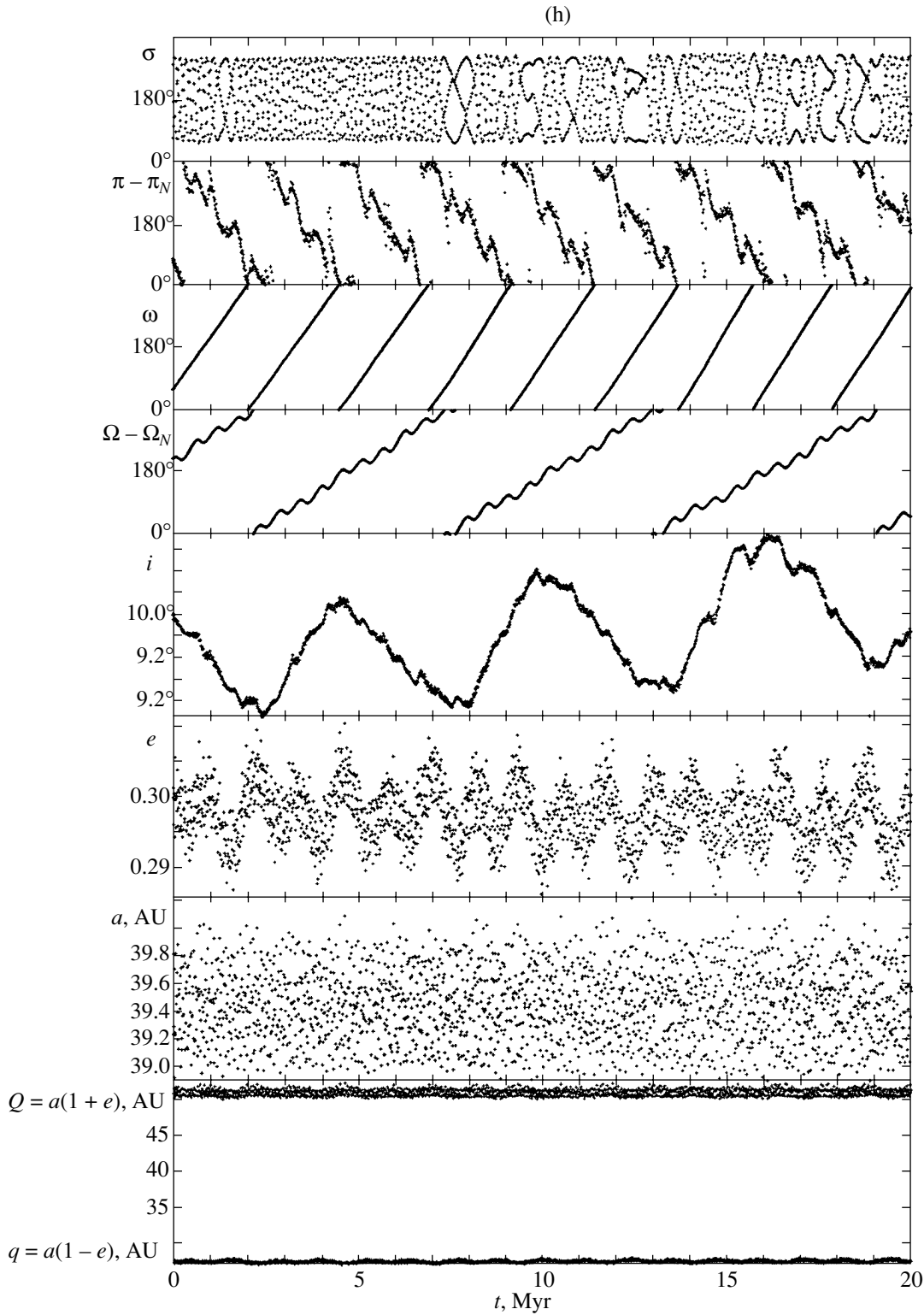

**Fig. 2.** Contd.

The amplitude of these variations in $i$ is usually smaller by a factor of 10 or more than the range of the main variations in $i$. Investigations of the three-body problem (the Sun, a planet, and a body) show (Lidov, 1961; Ipa-tov, 1992) that, in the case of the fixed orbit of a planet and for $a$ = const, we have max $i_r$ and min $e$ at $\omega = 0$ or $\omega = 180°$, and min $i_r$ and max $e$ at $\omega = \mp 90°$, where $i_r$ is the inclination of a body's orbit to the orbit of the planet

(i)

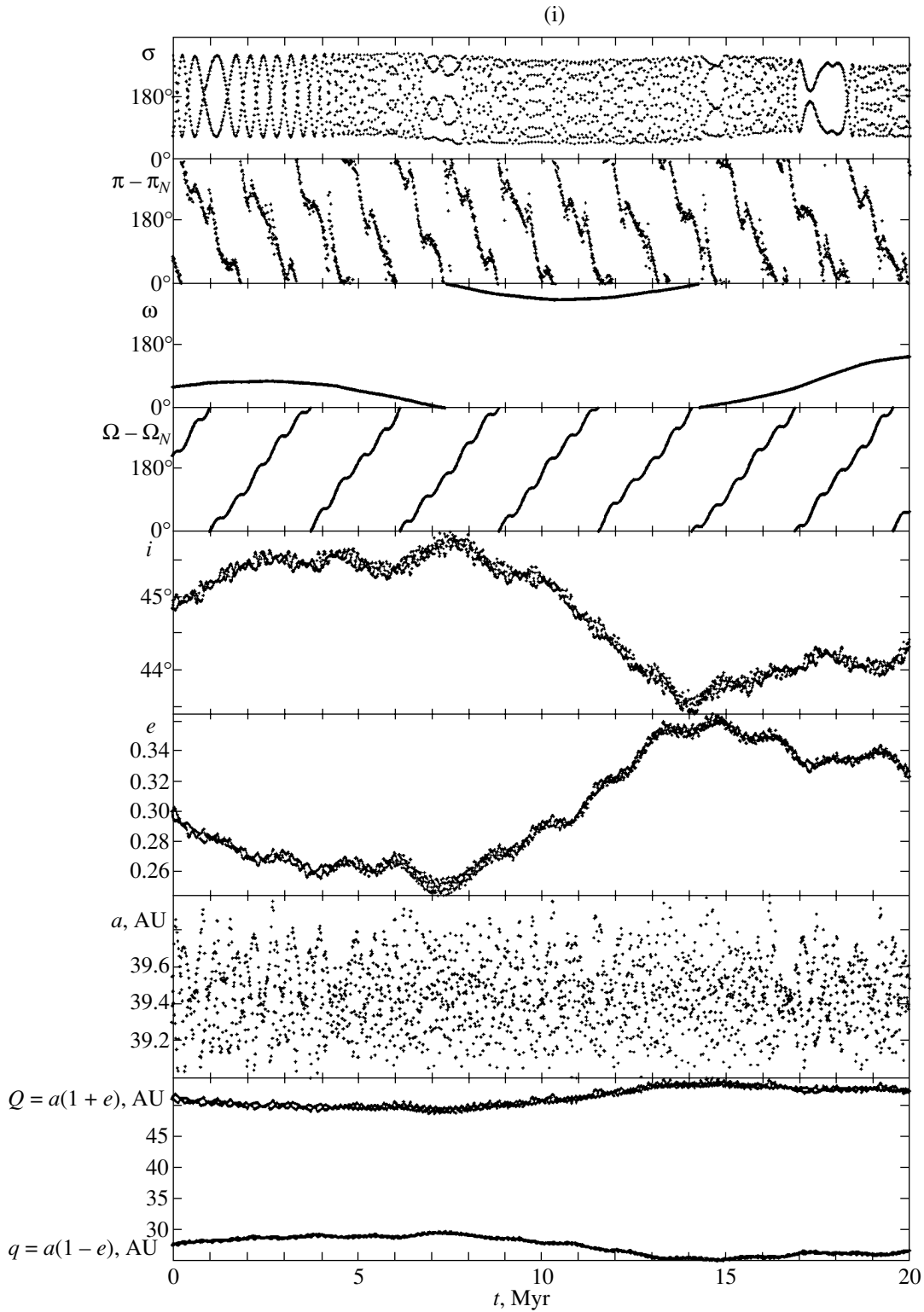

**Fig. 2.** Contd.

and $i_r < 90°$. Such relationships between the variations in $\omega$ and $i$ were obtained for all runs for the *DI* types (Fig. 2b) and *LD* types (Fig. 2f), and for some runs for the *ID* and *SI* types (Fig. 2d). For some runs for the types *ID* (Fig. 2a), *LI* (Fig. 2c) and *SI* we have min $i$ at $\omega = 0°$ or $\omega = 180°$. Note that for the same *ID* type at $\omega = 0°$ or $\omega = 180°$, we have min $i$ for large variations in the critical angle $\sigma$ and max $i$ for small (<180°) varia-tions in $\sigma$. For smaller values of $\sigma$, the intervals of variations in the orbital elements $a$, $e$, and $i$ are smaller.

For the 2 : 3 resonance with Neptune, $\Delta\pi = (\omega + \Omega) - (\omega_N + \Omega_N)$ usually decreases in the course of evolution (Figs. 2a, 2f, 2g). Sometimes $\Delta\pi$ mainly decreases (Fig. 2b). The $\nu_8$ resonance ($\Delta\pi \approx$ const) was usually obtained for the $LI$ type. In Fig. 2c, for $13 < t < 16.5$ Myr, $\Delta\pi$ varies between $160°$ and $360°$. For two other runs, $\Delta\pi$ varies in the ranges from $240°$–$360°$ and $180°$–$400°$ during 7 and 10 Myr, respectively. Note that, for the orbits located outside the mean motion resonances, a libration of $\Delta\pi$ was often obtained at $a_0 \approx 42$ AU (more rarely at $a_0 \approx 40$ AU), and for some runs it took place during the total time interval considered.

It is known that most of the observed trans-Neptunian objects with $a \leq 42$ AU are located near the 2 : 3 resonance with Neptune. In our runs, the orbits which had the $ID$ type with a small libration of $\sigma$ and were in this resonance, were the most stable. The orbits of the observed trans-Neptunian objects are not well known. In some cases, for small variations in the initial orbital elements, the type of the variations in $\Delta\Omega$ and $\omega$ can change and the limits of the variations in the orbital elements can differ widely. Therefore, one must consider the initial data in a certain vicinity of the initial orbital elements (the dimensions of this region depend on the errors in the determination of the orbits), when evaluating the lifetimes of actual trans-Neptunian bodies.

For all considered runs of the evolution of resonant orbits, the maximum values of $e$ and $i$ exceeded 0.07 and $3°$, respectively. The interval $\Delta a = a_{\max} - a_{\min}$ of the variations in $a$ for the 2 : 3 resonance with Neptune is about 1 AU. It may change by a factor of 1.8 for variations in $\Omega_0$, $\omega_0$, and $M_0$. For example, for $a_0 = 39.3$ AU, the $\Delta a$ value varied from 0.6 to 1.07 AU at $e_0 = 0.05$ and $i_0 = 5°$ and from 0.68 to 1.24 AU at $e_0 = 0.15$ and $i_0 = 5°$. For $\Omega_0 = \omega_0 = M_0 = 60°$ and $e_0 = 0.15$ took the values of about 1.1–1.2 AU at $0° \leq i_0 \leq 10°$ and 0.84–0.88 AU at $60° \leq i_0 \leq 90°$. In the case of $i_0 = 5°$, $\Delta a$ values increased from 0.86 to 1.24 AU as $e_0$ increases from 0 to 0.3. The above values of $\Delta a$ are presented for resonant orbits. If a body left the resonance, the values of $\Delta a$ could be much larger.

## CONCLUSION

The results of the numerical investigations showed that the character and the limits of the variations in orbital elements for the 2 : 3 resonance with Neptune can differ significantly for various initial orbital orientations and initial positions of bodies in orbits. If the difference $\Delta\Omega = \Omega - \Omega_N$ in the longitudes of the ascending node of the body and Neptune decreases and the argument of perihelion $\omega$ increases in the course of evolution, then most of the bodies leave the resonance in 20 Myr. In the case of an increase in $\Delta\Omega$ and a decrease in $\omega$, bodies remain in the resonance for a much longer time. The regions of eccentricities and inclinations, for which some bodies were in the secular resonance $\eta_{18}$ ($\Delta\Omega \approx$ const) and the Kozai resonance ($\omega \approx$ const), are larger than the regions predicted for small variations in a critical angle. Some bodies were in both these resonances at the same time.

## ACKNOWLEDGMENTS

For S.I. Ipatov, this work was supported by the Russian Foundation for Basic Research, project no. 96-02-17892, and by the Federal Scientific and Technical Program "Astronomy," section 1.9.4.1. The runs were made during the visit of S.I. Ipatov to Namur (Belgium) in 1995, supported by the ESO grant no. B-06-018.